\documentclass[12pt]{article}
\usepackage{epsf}

\begin{document}

\newcommand{\gtrsim}{ \mathop{}_{\textstyle \sim}^{\textstyle >} }
\newcommand{\lesssim}{ \mathop{}_{\textstyle \sim}^{\textstyle <} }

\newcommand{\rem}[1]{{\bf #1}}

\renewcommand{\theequation}{\thesection.\arabic{equation}}

\renewcommand{\thefootnote}{\fnsymbol{footnote}}
\setcounter{footnote}{0}
\begin{titlepage}

\def\thefootnote{\fnsymbol{footnote}}

\begin{center}

\hfill TU-769\\
\hfill May, 2006\\

\vskip .75in

{\Large \bf 
QCD Correction to Neutralino Annihilation Process \\
and Dark Matter Density in Supersymmetric Models
}

\vskip .75in

{\large
Takeo Moroi, Yukinari Sumino and Akira Yotsuyanagi
}

\vskip 0.25in

{\em
Department of Physics, Tohoku University,
Sendai 980-8578, JAPAN}

\end{center}
\vskip .5in

\begin{abstract}

We calculate QCD correction to the neutralino annihilation cross
section into quark anti-quark final state and discuss its implications
to the calculation of neutralino relic density.  We see that the QCD
correction enhances the pair-annihilation cross section by $O(10\ \%)$
when final-state quarks are non-relativistic.  Consequently, when the
lightest neutralinos dominantly annihilate into a $t\bar{t}$ pair, the
relic density of the lightest neutralino is significantly affected by
the QCD correction, in particular when the lightest-neutralino mass is
close to the top-quark mass.
  
\end{abstract}

\end{titlepage}

\renewcommand{\thepage}{\arabic{page}}
\setcounter{page}{1}
\renewcommand{\thefootnote}{\#\arabic{footnote}}
\setcounter{footnote}{0}

\section{Introduction}
\setcounter{equation}{0}
\label{sec:intro}

Low energy supersymmetry has been regarded as an attractive candidate
of new physics since it can solve various problems in the standard
model of particle physics.  One of the strong motivations of
supersymmetry is that it provides a plausible candidate of cold dark
matter.  Recent precise cosmological observations have provided strong
evidence of the existence of dark matter.  In particular, Wilkinson
Microwave Anisotropy Probe (WMAP) experiment has determined the
present density parameter of the dark matter, which is given by
\cite{WMAP2006}
\begin{eqnarray}
  \Omega_{\rm c} h_{100} ^2 = 0.105^{+0.007}_{-0.013}~~~
  (68\ \% \ {\rm C.L.}),
  \label{Omega_c(WMAP)}
\end{eqnarray}
where $h_{100}$ is Hubble constant in units of 100\ km/sec/Mpc.  From
the particle-physics point of view, however, this raises a serious
question since there is no viable candidate of dark matter in the
particle content of the standard model.  In supersymmetric models, the
lightest superparticle (LSP) is stable due to $R$-parity conservation
and hence it can be a candidate of cold dark matter.  It has been
widely recognized that, if the lightest neutralino $\chi^0_1$ is the
LSP, its thermal relic density can become consistent with the WMAP
value given in Eq.~(\ref{Omega_c(WMAP)}) in some parameter region.
For a better understanding of the evolution of the universe, it is
important to confirm or exclude the idea of neutralino (or LSP) dark
matter.

The first important step of such test is to find low-energy
supersymmetry in future collider experiments.  Once superparticles
are found, in addition, it is also important to quantitatively test
the idea of neutralino dark matter.  Quantitative test will be performed by
calculating the relic density of neutralino, using mass and coupling
parameters of superparticles measured in collider experiments
\cite{CDMatLC}.
Currently, cosmological observations have determined the dark matter
density with an accuracy of $O( 10\ \%)$, hence the
density parameter of the neutralino $\Omega_{\chi^0_1}$ is desired to
be calculated theoretically at the same (or better) accuracy.

As is well known, the relic density of neutralino is sensitive to the
neutralino pair annihilation cross section, which strongly depends on
mass and mixing parameters of the superparticles.\footnote{In some
case, coannihilation process also plays an important role.  In this
paper, however, we assume that the mass of the lightest neutralino is
quite far from the second-to-the-lightest superparticle, and that the
coannihilation process is ineffective.} Thus, for a precise
calculation of the relic density, detailed calculation of the pair
annihilation cross section is necessary.  Indeed, in the case where
the lightest neutralino is the LSP, the pair annihilation cross
section and relic density of the lightest neutralino have been
calculated in detail, in particular in connection with the WMAP
observation \cite{LSPDM_WMAP}.

So far, most of the previous calculations of pair annihilation cross
section have been based on tree-level calculation.  (See, however,
\cite{Barger:2005ve}.)  One may naively
estimate the size of one-loop
corrections to be $O(\lambda^2/16\pi^2)$ with $\lambda$ being relevant
coupling constant, hence it is $O(1\ \%)$ or smaller as far as
$\lambda\lesssim 1$.  In such a case, radiative corrections to the pair
annihilation cross section may be unimportant for the purpose of
calculating $\Omega_{\chi^0_1}$ with an accuracy of $O( 10\ \%)$.
If we consider QCD correction, however, this is not always the case.
If we consider processes with $q\bar{q}$ final state (with $q$ and
$\bar{q}$ being quark and anti-quark, respectively), QCD correction
becomes $O(10\ \%)$ when final-state quarks are non-relativistic.  In
the so-called ``focus-point'' model \cite{focus}, for example,
the annihilation process into a top quark pair, 
$\chi^0_1\chi^0_1\rightarrow
t\bar{t}$, becomes one of the dominant pair annihilation processes
\cite{Feng:2000gh,Ibe:2005jf}.  Then, QCD correction becomes important
when the lightest neutralino mass is close to the top-quark mass
due to threshold enhancement of the QCD correction.
(The threshold enhancement 
is well known e.g.\ from the analysis of the cross section
for $e^+e^- \to t\bar{t}$
close to $t\bar{t}$ threshold \cite{Hoang:2000yr}).

The main purpose of this paper is to calculate the next-to-leading order
(NLO) QCD correction to the neutralino annihilation
cross section into $q\bar{q}$ final state
and to discuss its implications to the calculation of the relic density
of the lightest neutralino.  (In fact, we also
have to consider real gluon emission to take care of 
infrared divergences.  Thus, we consider processes of the type 
$\chi^0_1\chi^0_1\rightarrow q\bar{q}(g)$, with $g$ being the gluon.)  As
we will see, QCD correction may become $O(10\ \%)$.  Since the relic
density of the lightest neutralino is approximately inversely
proportional to pair annihilation cross section, this has certain
significance for a precise calculation of the dark matter density.  

In addition, for scenarios where one of very weakly interacting
superparticles (like gravitino \cite{GravitinoCDM}, axino
\cite{AxinoCDM}, or right-handed sneutrino \cite{Asaka:2005cn}, which
are recently called superWIMPs) is the LSP, study of the QCD
correction has some significance.  In such scenario, the lightest
neutralino becomes unstable and decays into a superWIMP (and other
standard-model particles), assuming that it is the lightest
superparticle in the minimal-supersymmetric-standard-model (MSSM)
sector.  Then, cosmologically, the lightest neutralino becomes an
important source of the superWIMP, which is now a candidate of cold
dark matter.  The important point is that the lifetime of the lightest
neutralino is very long; in the thermal history of the universe, the
decay occurs after the lightest neutralino freezes out from the
thermal bath.  In this case, abundance of the superWIMP produced by
the decay of the lightest neutralino is given by the freeze-out
density of the lightest neutralino.  Thus, even if one of the
superWIMPs is the LSP, precise determination of the freeze-out density
of the lightest neutralino is also important.\footnote{
One of the authors resisted glorifying supersymmetric models as
scenarios for new physics, but the other authors ganged up and forced
through the present style of Introduction.}

This paper is organized as follows.  In the next section, we summarize
the framework of our study.  Then, in Section \ref{sec:QCDcorrection},
we present formulas for QCD correction to the process
$\chi^0_1\chi^0_1\rightarrow q\bar{q}(g)$.  In Section
\ref{sec:Omega}, we calculate the relic density of the lightest
neutralino and discuss implications of the QCD correction to the
calculation of the dark matter density.  Section \ref{sec:conclusion} is
devoted to conclusions and discussion.

\section{Framework}
\setcounter{equation}{0}
\label{sec:framework}

In our study, we work in the framework of the MSSM.  In addition, we
assume that the lightest neutralino $\chi^0_1$ becomes the LSP and
hence is stable.  In general, the mass and mixing parameters of
superparticles depend on various MSSM parameters.  To make our points
clearer, we adopt several simplifications.

In calculating $\Omega_{\chi^0_1}$, we need to fix properties of the
lightest neutralino $\chi^0_1$, which are determined from the
neutralino mass matrix ${\cal M}_{\chi^0}$.  Denoting the $U(1)_Y$ and
$SU(2)_L$ gaugino mass parameters as $m_{\rm G1}$ and $m_{\rm G2}$ and
supersymmetric Higgs mass as $\mu_H$, ${\cal M}_{\chi^0}$ is given by
\begin{eqnarray}
  {\cal M}_{\chi^0} = \left( \begin{array}{cccc}
    -m_{\rm G1} & 0 & - g_1 v \cos\beta & g_1 v \sin\beta \\
    0 & -m_{\rm G2} & g_2 v \cos\beta & - g_2 v \sin\beta \\
    - g_1 v \cos\beta & g_2 v \cos\beta & 0 & \mu_H \\
    g_1 v \sin\beta & - g_2 v \sin\beta & \mu_H & 0
  \end{array} \right),
\end{eqnarray}
where $g_1$ and $g_2$ are gauge coupling constants for for $U(1)_Y$
and $SU(2)_L$ gauge groups, respectively, $v\simeq 174\ {\rm GeV}$ is
the (total) vacuum expectation value of the Higgs boson, and $\tan\beta$
is the ratio of the vacuum expectation values of the up- and down-type Higgs
bosons.  This mass matrix is diagonalized by a unitary matrix
$U_{\chi^0}$ as
\begin{eqnarray}
  U_{\chi^0}^T {\cal M}_{\chi^0} U_{\chi^0}
  = {\rm diag}
  (m_{\chi^0_1}, m_{\chi^0_2}, m_{\chi^0_3}, m_{\chi^0_4}),
\end{eqnarray}
and the lightest neutralino is given by
\begin{eqnarray}
  \chi^0_1 = [U_{\chi^0}]_{11}^* \tilde{B} 
  + [U_{\chi^0}]_{21}^* \tilde{W}^0
  + [U_{\chi^0}]_{31}^* \tilde{H}^0_d
  + [U_{\chi^0}]_{41}^* \tilde{H}^0_u,
\end{eqnarray}
where $\tilde{B}$, $\tilde{W}^0$, $\tilde{H}^0_u$, and $\tilde{H}^0_d$
are Bino, (neutral) Wino, up-type Higgsino, and down-type Higgsino,
respectively.  

In our study, we adopt the grand-unified-theory (GUT) relation among
gaugino masses.
In this case we obtain the relation between $U(1)_Y$ and
$SU(2)_L$ gaugino masses, $m_{\rm G1}$ and $m_{\rm G2}$, as
\begin{eqnarray}
  \frac{m_{\rm G2}}{g_2^2} = 
  \frac{3}{5}\frac{m_{\rm G1}}{g_1^2}.
\end{eqnarray}
Then, since $m_{\rm G2}>m_{\rm G1}$, the lightest neutralino is
approximately given by a linear combination of Bino and Higgsinos.  We
will see that the QCD correction to the process
$\chi^0_1\chi^0_1\rightarrow t\bar{t}(g)$ becomes most significant
when the lightest neutralino is mostly Bino-like but is with a sizable
Higgsino contamination.  Notice that the formulas for the QCD
correction which will be given in the next section are independent of
the relation among the gauginos.

With the unitary matrix $U_{\chi^0}$, the $Z$-$\chi^0_1$-$\chi^0_1$
vertex, which is relevant for the process
$\chi^0_1\chi^0_1\rightarrow q\bar{q}(g)$, is given in the form
\begin{eqnarray}
  {\cal L}_{Z\chi^0\chi^0} = 
  \frac{1}{2} A_\chi \bar{\chi}^0_1 \gamma_\mu \gamma_5 \chi^0_1 Z_\mu,
\end{eqnarray}
where 
\begin{eqnarray}
  A_\chi = \frac{1}{2} g_z 
  \left( \left[U_{\chi^0}\right]^*_{31} \left[U_{\chi^0}\right]_{31}
    - \left[U_{\chi^0}\right]^*_{41} \left[U_{\chi^0}\right]_{41}
  \right),
  \label{A_chi}
\end{eqnarray}
with $g_Z\equiv \sqrt{g_1^2+g_2^2}$.  When $m_{\rm G1}\ll\mu_H$,
$[U_{\chi^0}]_{11}\simeq 1$ while $[U_{\chi^0}]_{i1}$ ($i=2,3,4$)
become close to $0$.  In this case, the lightest neutralino is almost
Bino and its interactions with gauge and Higgs bosons are suppressed.
Consequently, dominant annihilation processes are into lepton pairs
(as far as the sleptons are lighter than the squarks) and hence are
$p$-wave processes.  When $\mu_H$ is relatively small, on the
contrary, Higgsino component in the lightest neutralino becomes
enhanced and $[U_{\chi^0}]_{31}$ and $[U_{\chi^0}]_{41}$ become
sizable.  In this case, the lightest neutralino pair dominantly
annihilates into $t\bar{t}$, $W^+W^-$, and $ZZ$ final states since
they are $s$-wave processes.

We also present the interaction of the neutralino with the lightest
(standard-model-like) Higgs boson $h$.  The interaction term is given
in the form
\begin{eqnarray}
  {\cal L}_{h\chi^0\chi^0} = 
  \frac{1}{2} Y_\chi \bar{\chi}^0_1 \chi^0_1 h.
\end{eqnarray}
In our following study, we consider the so-called decoupling limit
where all the physical Higgs bosons except the standard-model-like
Higgs $h$ are very heavy.  In this case, we can neglect effects of
heavier Higgses on the annihilation processes of the lightest
neutralino.  In addition, in the decoupling limit, $Y_\chi$ is well
approximated by\footnote{ In the general case, $Y_\chi$ is given by
replacing $\beta\rightarrow\alpha +\frac{1}{2}\pi$, where $\alpha$ is
the mixing angle in the CP-even Higgs sector.  (See, for example,
\cite{Gunion:1989we}.)  }
\begin{eqnarray}
  Y_\chi = - \left( g_2 \left[U_{\chi^0}\right]_{21} - 
    g_1 \left[U_{\chi^0}\right]_{11} \right)
  \left( \left[U_{\chi^0}\right]_{31} \cos\beta -
    \left[U_{\chi^0}\right]_{41} \sin\beta \right).
\end{eqnarray}
(Here and hereafter, we work in the bases where the matrix
$U_{\chi^0}$ is real.)

In the calculation of $\Omega_{\chi^0_1}$, sfermions may also
contribute.  In the following, we pay particular attention to the case
where the annihilation process of $\chi^0_1$ is dominated by $s$-wave
processes.  In such a case, sfermion-exchange diagrams are mostly
subdominant since they induce $p$-wave processes.  One exception may
be $t$-channel exchange of stops.  We assume that the stops are heavy
and their contribution to the pair annihilation of $\chi^0_1$ is
negligible.  In addition, we also assume that heavier Higgs bosons are
heavy enough so that they do not significantly affect
$\Omega_{\chi^0_1}$.  Such a mass spectrum with heavy scalars is
realized in, for example, focus-point models \cite{focus}.

\section{QCD Correction to $\sigma_{\chi^0_1\chi^0_1\rightarrow
    q\bar{q}(g)}$}
\setcounter{equation}{0}
\label{sec:QCDcorrection}

We calculate the NLO QCD correction to the cross section for
neutralino pair annihilation into $q\bar{q}(g)$ final state.  In this
paper, we consider the case where the squarks (in particular, stops)
are much heavier than the $Z$-boson.  This is the case in a large
class of supersymmetric models, partially due to renormalization group
effects via gluino mass.  In such models, the decays
$\chi^0_1\chi^0_1\rightarrow q\bar{q}(g)$ are induced dominantly via
$Z$ and $h$-boson exchange diagrams.  When the
initial-state neutralinos are non-relativistic, $Z$-boson exchange
diagram turns out to be most important.  $Z$ and $h$-boson exchange
contributions do not interfere, since the $P$ (parity) properties of
$Z$ and $h$ are different; hence we may discuss their contributions
separately.

\subsection{$O(\alpha_s)$ correction}

We first consider $O(\alpha_s)$ contributions from the one-loop and
real-gluon emission processes.  As will be discussed below, QCD
corrections are enhanced when the final-state quarks are
non-relativistic due to boundstate effects.  In such a case, it is
necessary to take account of (some part of) higher-order corrections
in order to obtain reliable predictions.  This will be treated in the
next subsection.

We start with the contribution of the $Z$-boson exchange diagrams.
The relevant interaction of the $Z$-boson with quarks can be
parameterized as
\begin{eqnarray}
  {\cal L}_{Zq\bar{q}} = 
  \sum_q \bar{q} \gamma_\mu (V_q + A_q \gamma_5) q Z_\mu.
\end{eqnarray}
As we will see, in the calculation of the relic density of $\chi^0_1$,
$t\bar{t}(g)$ final state is of particular interest.  For the top quark,
\begin{eqnarray}
  V_t = \frac{1}{2g_Z} 
  \left( \frac{1}{2} g_2^2 - \frac{5}{6} g_1^2 \right),~~~
  A_t = - \frac{1}{4} g_Z.
\end{eqnarray}

\begin{figure}[t]
  \centerline{\epsfxsize=0.6\textwidth\epsfbox{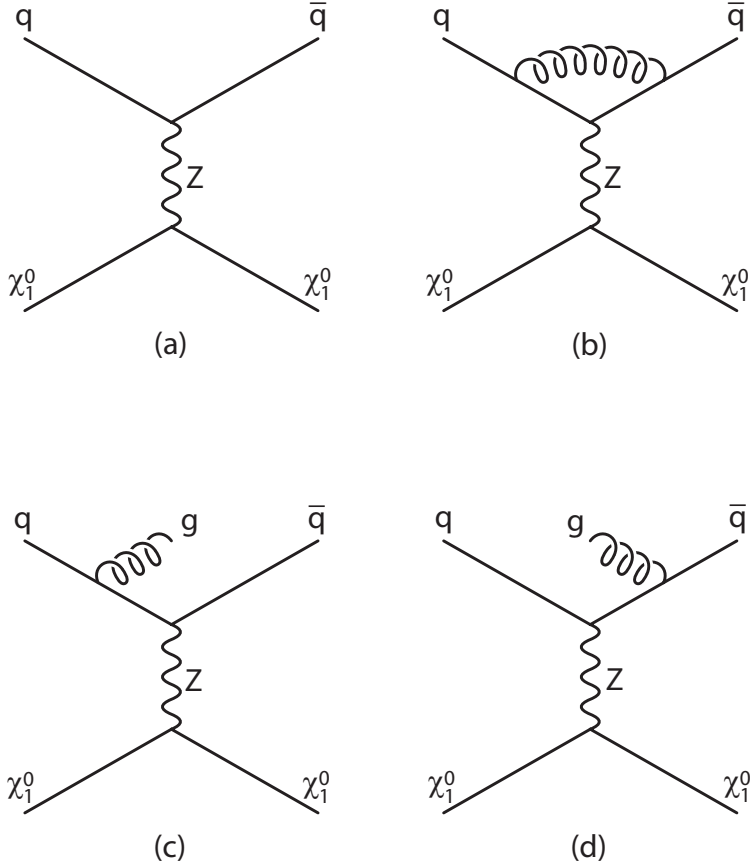}}
  \caption{Feynman diagrams for the processes
    $\chi^0_1\chi^0_1\rightarrow Z^*\rightarrow q\bar{q}$ ((a) and
    (b)) and $\chi^0_1\chi^0_1\rightarrow Z^*\rightarrow q\bar{q}g$
    ((c) and (d)).}
  \label{fig:feyndiag}
\end{figure}

We should consider two contributions simultaneously; one is the
virtual correction to $Z$-$q$-$\bar{q}$ vertex and the other is the
real gluon emission process $\chi^0_1\chi^0_1\rightarrow
Z^*\rightarrow q\bar{q}g$.  Both of them, individually, are infrared
divergent, but these divergences cancel when both contributions are
summed.  Let us denote the Feynman amplitude for the process
$\chi^0_1\chi^0_1\rightarrow Z^*\rightarrow q\bar{q}(g)$ as
\begin{eqnarray}
  {\cal M}_{\chi^0_1\chi^0_1\rightarrow Z^*\rightarrow q\bar{q}(g)} = 
  \Lambda^{(\chi)}_\mu D_{\mu\nu}^{(Z)} \Lambda^{(q)}_\nu.
\end{eqnarray}
Here, $D_{\mu\nu}^{(Z)}=\frac{1}{Q^2-m_Z^2}(g_{\mu\nu}-\frac{Q_\mu
Q_\nu}{m_Z^2})$ is the propagator of $Z$-boson (with $Q_\mu$ being the
total four-momentum of the system), while $\Lambda^{(\chi)}_\mu$ and
$\Lambda^{(q)}_\nu$ are bi-spinors (including coupling constants)
consisting of the wave functions of the neutralino and quark,
respectively.  It is convenient to define the following quantity:
\begin{eqnarray}
  \Pi^{(q)}_{\mu\nu} = \sum_{\rm spin} 
  \int d\Phi_{q\bar{q}} \left[ \Lambda^{(q)}_\mu \Lambda^{(q)*}_\nu 
  \right]_{\chi^0_1\chi^0_1\rightarrow Z^*\rightarrow q\bar{q}} +
  \sum_{\rm spin}
  \int d\Phi_{q\bar{q}g} \left[ \Lambda^{(q)}_\mu \Lambda^{(q)*}_\nu 
  \right]_{\chi^0_1\chi^0_1\rightarrow Z^*\rightarrow q\bar{q}g},
  \label{Pi_mn}
\end{eqnarray}
where the first and second terms represent contributions from the
processes $\chi^0_1\chi^0_1\rightarrow Z^*\rightarrow q\bar{q}$ (i.e.,
leading order $+$ virtual correction) and $\chi^0_1\chi^0_1\rightarrow
Z^*\rightarrow q\bar{q}g$ (i.e., real emission), respectively.  In the
above formula, the integrals $\int d\Phi$ are carried out over the
final-state phase space, and the spin sums of the final-state
particles are taken.  The first and second terms in Eq.\ (\ref{Pi_mn})
individually contain infrared divergences, while $\Pi^{(q)}_{\mu\nu}$
is finite.  We decompose $\Pi^{(q)}_{\mu\nu}$ into the vector and
scalar parts as
\begin{eqnarray}
  \Pi^{(q)}_{\mu\nu} = 
  \Pi^{(q)}_{\rm V} (Q^2)
  \left( -g_{\mu\nu} + \frac{Q_\mu Q_\nu}{Q^2} \right) 
  + \Pi^{(q)}_{\rm S} (Q^2) \frac{Q_\mu Q_\nu}{Q^2} .
\end{eqnarray}

For the purpose of the calculation of the relic density of neutralino,
we define\footnote{It is often the case that the quantity ${\cal S}$
is expressed as $v_{\rm rel}\sigma$, where $v_{\rm rel}$ is the
relative velocity of the initial-state neutralinos and $\sigma$ is the
neutralino pair-annihilation cross section.}
\begin{eqnarray}
  {\cal S}_{\chi^0_1\chi^0_1\rightarrow Z^*\rightarrow q\bar{q}(g)} (Q^2) &=&
  2 \, v_{\chi^0_1} \,
  \sigma_{\chi^0_1\chi^0_1\rightarrow Z^*\rightarrow q\bar{q}(g)}
  \nonumber \\ &=&
  \frac{1}{4Q^2} 
  \left[ 
    \sum_{\rm spin} 
    \int d\Phi_{q\bar{q}} 
    \left| {\cal M}_{\chi^0_1\chi^0_1\rightarrow Z^*\rightarrow q\bar{q}} 
    \right|^2
    + \sum_{\rm spin}
    \int d\Phi_{q\bar{q}g} \left| 
      {\cal M}_{\chi^0_1\chi^0_1\rightarrow Z^*\rightarrow q\bar{q}g}
    \right|^2
  \right],
  \nonumber \\
\end{eqnarray}
where $v_{\chi^0_1}$ is the velocity of the lightest neutralino in the
center-of-mass frame, which is given by
\begin{eqnarray}
  v_{\chi^0_1}=\sqrt{1-\frac{4 m_{\chi^0_1}^2}{Q^2}}.
\end{eqnarray}
In addition, $\sigma_{\chi^0_1\chi^0\rightarrow Z^*\rightarrow
q\bar{q}(g)}$ is the total cross section for the process
$\chi^0_1\chi^0\rightarrow Z^*\rightarrow q\bar{q}(g)$, hence, in the
final formula, the spins of neutralinos are understood to be averaged.
As we will see, the function ${\cal S}_{\chi^0_1\chi^0\rightarrow
Z^*\rightarrow q\bar{q}(g)}$ plays an important role in the
calculation of $\Omega_{\chi^0_1}$.  With $\Pi^{(q)}_{\rm V}$ and
$\Pi^{(q)}_{\rm S}$, ${\cal S}_{\chi^0_1\chi^0\rightarrow
Z^*\rightarrow q\bar{q}(g)}$ is expressed as
\begin{eqnarray}
  {\cal S}_{\chi^0_1\chi^0\rightarrow Z^*\rightarrow q\bar{q}(g)} (Q^2)
  =
  \frac{1}{Q^2} \left[
    \frac{Q^2-4 m_{\chi^0_1}^2}{(Q^2-m_Z^2)^2} \Pi^{(q)}_{\rm V} (Q^2)
    + \frac{2 m_{\chi^0_1}^2}{m_Z^4} \Pi^{(q)}_{\rm S} (Q^2)
  \right] \left| A_\chi \right|^2.
\end{eqnarray}
Importantly, the first term in the parenthesis is proportional to
$Q^2-4m_{\chi^0_1}^2$; as a result, it is suppressed when the initial
state neutralinos are non-relativistic.  Thus, in such a case, the
contribution of $\Pi^{(q)}_{\rm V}$ becomes negligible.  In addition,
$\Pi^{(q)}_{\rm S}$ is proportional to the mass-squared of the
final-state quark.  Since the relic density is determined by the cross
section at $Q^2\sim 4m_{\chi^0_1}^2$, we pay particular attention to
the process $\chi^0_1\chi^0_1\rightarrow Z^*\rightarrow t\bar{t}(g)$
and neglect other processes concerning lighter quarks, in calculating
the relic density of $\chi^0_1$.

In this subsection, we present the formulas for $\Pi^{(q)}_{\rm V}$
and $\Pi^{(q)}_{\rm A}$ up to $O(\alpha_s)$.  Corresponding Feynman
diagrams are shown in Fig.\ \ref{fig:feyndiag}.  At the tree level, we
obtain\footnote{At $O(\alpha_s)$, the correction from real gluon
emission is also from tree diagrams.  $[\Pi^{(q)}_{\rm V}]_{\rm tree}$
and $[\Pi^{(q)}_{\rm S}]_{\rm tree}$ do not include such
contributions.}
\begin{eqnarray}
  \left[ \Pi^{(q)}_{\rm V}  (Q^2) \right]_{\rm tree} &=& 
  \frac{m_q^2}{3\pi (1-v_q^2)} 
  v_q \left[ (3-v_q^2) \left| V_q \right|^2 +
    2 v_q^2 \left| A_q \right|^2 
  \right],
  \\
  \left[ \Pi^{(q)}_{\rm S} (Q^2) \right]_{\rm tree} &=& 
  \frac{m_q^2}{\pi} v_q \left| A_q \right|^2,
\end{eqnarray}
where
\begin{eqnarray}
v_q = \sqrt{1 - \frac{4 m_q^2}{Q^2}}.
\end{eqnarray}

All information on the QCD correction (at $O(\alpha_s)$) is
contained in $\Pi^{(q)}_{\rm V}$ and $\Pi^{(q)}_{\rm S}$.  In the
calculation of the virtual contribution (Fig.\ \ref{fig:feyndiag}(b)),
we adopt the on-mass-shell renormalization condition to determine the
vertex counter term: we require that the $O(\alpha_s)$ correction to the
vector vertex (which is proportional to $V_q$) vanishes when
$Q^2\rightarrow 0$.  Then, $O(\alpha_s)$ correction to $\Pi^{(q)}_{\rm
  V}$ is found to be \cite{Hikasa:1996wp}
\begin{eqnarray}
  \left[ \Pi^{(q)}_{\rm V} (Q^2) \right]_{O(\alpha_s)} &=& 
  \frac{\alpha_sC_2 m_q^2}{24\pi^2(1-v_q^2)} \left|V_q\right|^2
  \Bigg[ 8 (3 - v_q^2) A(v_q) + 
  6 v_q ( 5 - 3 v_q^2)
  \nonumber \\ &&
  + ( 33 + 22 v_q^2 - 7 v_q^4 )
  \ln \frac{1+v_q}{1-v_q}
  \Bigg]
  \nonumber \\ &&
  + \frac{\alpha_sC_2 m_q^2}{48\pi^2(1-v_q^2)} \left|A_q\right|^2
  \Bigg[ 32 v_q^2 A(v_q) +
  6 v_q (-7 + 10 v_q^2 + v_q^4)
  \nonumber \\ &&
  + ( 21 + 59 v_q^2 + 19 v_q^4 - 3 v_q^6 )
  \ln \frac{1+v_q}{1-v_q}
  \Bigg],
\end{eqnarray}
where $C_2=\frac{4}{3}$ is quadratic Casimir operator for the
fundamental representation of $SU(3)$, 
\begin{eqnarray}
  A(v_q) &=& (1+v_q^2) \Bigg[ 
  4 {\rm Li}_2 \left(\frac{1-v_q}{1+v_q}\right)
  + 2 {\rm Li}_2 \left(-\frac{1-v_q}{1+v_q}\right)
  - 3 \ln \frac{2}{1+v_q} \ln \frac{1+v_q}{1-v_q}
  \nonumber \\ &&
  - 2 \ln v_q \ln \frac{1+v_q}{1-v_q} \Bigg]
  - 3 v_q \ln \frac{4}{1-v_q^2}
  - 4 v_q \ln v_q,
\end{eqnarray}
and ${\rm Li}_2(x)$ is the dilogarithm (Spence) function
\begin{eqnarray}
  {\rm Li}_2 (x) = - \int_0^x dt \frac{\ln (1-t)}{t}.
\end{eqnarray}
In addition, $O(\alpha_s)$ correction to $\Pi^{(q)}_{\rm S}$ is
given by
\begin{eqnarray}
  \left[ \Pi^{(q)}_{\rm S} (Q^2) \right]_{O(\alpha_s)} =
  \frac{\alpha_sC_2 m_q^2}{16\pi^2} \left|A_q\right|^2
  \Bigg[ 16 A(v_q)
  + 6 v_q (7 - v_q^2)
  + ( 19 + 2 v_q^2 + 3 v_q^4 )
  \ln \frac{1+v_q}{1-v_q}
  \Bigg].
  \nonumber \\
\end{eqnarray}
We have checked that the ratio $[\Pi^{(q)}_{\rm S}
(Q^2)]_{O(\alpha_s)}/[\Pi^{(q)}_{\rm S} (Q^2)]_{\rm tree}$ coincides with
the correction factor for the decay rate of the pseudo-scalar Higgs boson
given in \cite{Drees:1990dq}.

We can also calculate the $h$-exchange contributions.  Although the
process $\chi^0_1\chi^0_1\rightarrow h^*\rightarrow t\bar{t}(g)$ is
$p$-wave suppressed and hence is subdominant, we write down the
formula for the QCD correction to this process to make this paper
self-contained.  Let us denote the $hq\bar{q}$ vertex as
\begin{eqnarray}
  {\cal L}_{hq\bar{q}} = 
  \sum_q Y_q \bar{q} q h,
\end{eqnarray}
and define
\begin{eqnarray}
  {\cal S}_{\chi^0_1\chi^0_1\rightarrow h^*\rightarrow q\bar{q}(g)} (Q^2) =
  2 \, v_{\chi^0_1} \, 
  \sigma_{\chi^0_1\chi^0_1\rightarrow h^*\rightarrow q\bar{q}(g)}.
\end{eqnarray}
Adopting on-mass-shell renormalization condition for the
$h$-$q$-$\bar{q}$ vertex, the coupling constant $Y_q$ is related to
the (on-shell) mass of $q$ as
\begin{eqnarray}
  Y_q = \frac{m_q}{\sqrt{2}v}.
\end{eqnarray}
Then, ${\cal S}_{\chi^0_1\chi^0_1\rightarrow h^*\rightarrow
  q\bar{q}(g)}$ is given by
\begin{eqnarray}
  {\cal S}_{\chi^0_1\chi^0_1\rightarrow h^*\rightarrow q\bar{q}(g)} (Q^2) =
  \frac{m_q^2(Q^2-4m_{\chi^0_1}^2)}{8\pi Q^2(Q^2-m_h^2)^2}
  v_q^3 \left| Y_q Y_\chi \right|^2
  \left( 1 + \frac{C_2\alpha_s}{\pi} \Delta_h
  \right),
\end{eqnarray}
where \cite{Drees:1990dq}
\begin{eqnarray}
  \Delta_h = \frac{1}{v_q} A(v_q) 
  + \frac{1}{16v_q^3} (3 + 34 v_q^2 - 13v_q^4) \ln \frac{1+v_q}{1-v_q}
  + \frac{3}{8v_q^2} (-1 + 7v_q^2).
\end{eqnarray}

For the calculation of the relic density of neutralino,  the $s$-wave
contribution plays the most significant role.  As we mentioned, for the
$s$-wave process, $t\bar{t}(g)$ final state is most important
while the effects of lighter quarks are unimportant (if we restrict
ourselves to the $Z$-exchange diagrams).  Thus, in order to see the
size of the QCD correction to the $s$-wave part, we define
\begin{eqnarray}
  R_{O(\alpha_s)} (Q^2) = 
  \frac{[ \Pi_{\rm S}^{(t)} (Q^2)  ]_{O(\alpha_s)}}
  {[ \Pi_{\rm S}^{(t)} (Q^2)  ]_{\rm tree}}.
\end{eqnarray}
We can see that the $O(\alpha_s)$ correction to the $s$-wave part is
enhanced when $v_t$ becomes smaller.  Indeed, expanding
$R_{O(\alpha_s)}$ around $v_t=0$, we obtain
\begin{eqnarray}
  R_{O(\alpha_s)} =
  \frac{C_2 \alpha_s}{\pi} 
  \left[ \frac{\pi^2}{2 v_t} -3 + O(v_t)\right].
  \label{expinvq}
\end{eqnarray}
This ratio is inversely proportional to $v_t$ when $v_t\ll 1$.  As a
result, the QCD correction largely enhances the neutralino
annihilation cross section especially when the final-state top and
anti-top quarks are non-relativistic.  Thus, in the calculation of the
relic density of the lightest neutralino, QCD correction may become
important.

In fact, it is well known that QCD corrections are enhanced as $v_t$
becomes smaller, due to boundstate effects.  Naive perturbative
expansion breaks down in the region $v_t \lesssim \alpha_s$.  In this
region, systematic treatment of boundstate effects are known for heavy
quarks, such as top quark.  Up to NLO, this corresponds (formally) to
the resummation of terms of the form $(\alpha_s/v_t)^k$ and
$v_t(\alpha_s/v_t)^k=\alpha_s(\alpha_s/v_t)^{k-1}$ for $0 \leq k <
\infty$ (where $k$ is an integer).  In the next subsection, we compute
the QCD correction reliably by resumming all these corrections.

\subsection{NLO annihilation cross section close to $q\bar{q}$ threshold}

We compute the QCD correction to the process
$\chi^0_1\chi^0_1\rightarrow q\bar{q}(g)$ close to $q\bar{q}$
threshold up to NLO (including boundstate effects).  Our main concern
is to calculate the relic density of the lightest neutralino; for this
purpose, the $s$-wave part of the cross section is most important.
Thus, in this subsection, we concentrate on the QCD correction to the
$s$-wave part.

Following the standard framework developed for the threshold cross
sections, for example, for the process $e^+e^-\rightarrow t\bar{t}$
\cite{Hoang:2000yr}, the NLO annihilation cross section (when the
final-state quarks are non-relativistic) is given as follows:
\begin{eqnarray}
  \left[ \Pi^{(q)}_{\rm S} (Q^2) \right]_{\rm NLO,\, NR} =
  F(Q^2)\,
  \left( 1 + \frac{C_2 \alpha_s}{\pi} \Delta_{\rm S}^{\rm (hard)}
  \right)
  \left[ \Pi^{(q)}_{\rm S} (Q^2) \right]_{\rm tree},
  \label{Pi(resum)}
\end{eqnarray}
where $\Delta_{\rm S}^{\rm (hard)}$ represents the hard-vertex
correction given by
\begin{eqnarray}
  \Delta_{\rm S}^{\rm (hard)} = \lim_{v_q \to 0}
  \left[
  \left( \frac{C_2 \alpha_s}{\pi} \right)^{-1}
  \frac{[ \Pi^{(q)}_{\rm S} (Q^2) ]_{O(\alpha_s)}}
  {[ \Pi^{(q)}_{\rm S} (Q^2) ]_{\rm tree}}
  - \frac{\pi^2}{2 v_q} 
  \right] \, = \, -3.
\end{eqnarray}
Boundstate effects are included in $F(Q^2)$, which is given in terms
of the Green functions of the non-relativistic Schr\"odinger equation:
\begin{eqnarray}
  F(Q^2)=\frac{{\rm Im}[G(\vec{0};E)]}{{\rm Im}[G_0(\vec{0};E)]} .
\end{eqnarray}
$G(\vec{x};E)$ is defined by
\begin{eqnarray}
 \left[
 (E+i\Gamma_q)-
 \left\{ -\frac{\nabla^2}{m_q} + V_{\rm QCD}^{\rm (NLO)}(r;\mu_{\rm Bohr})
 \right\}
 \right] G(\vec{x};E) = \delta^3(\vec{x}) .
\label{schroedingereq}
\end{eqnarray}
Here, $V_{\rm QCD}^{\rm (NLO)}(r;\mu_{\rm Bohr})$ denotes the static
QCD potential up to NLO \cite{Fischler:1977yf}; $E=\sqrt{Q^2}-2m_q$ is
the center-of-mass energy measured from the $q\bar{q}$ threshold, and
$\Gamma_q$ is the total decay width of $q$.  On the other hand,
$G_0(\vec{x};E)$ is the non-relativistic Green function of a free
$q\bar{q}$ pair, which is defined via Eq.\ (\ref{schroedingereq})
after setting $V_{\rm QCD}^{\rm (NLO)}(r)$ to $0$.

In the case of QED (i.e., when $V_{\rm QCD}(r)$ is replaced by the
Coulomb potential $-\alpha_{\rm QED}/r$, with $\alpha_{\rm QED}$ being
the fine-structure constant), $F(Q^2)$ reduces to the well-known
Sommerfeld factor:
\begin{eqnarray}
    \left[ F(Q^2) \right]_{\rm QED} =
    \frac{z}{1-e^{-z}},
    \label{Sommerfeldfac}
\end{eqnarray}
where
\begin{eqnarray}
    z = \frac{\pi \, \alpha_{\rm QED}}{v_q} .
\end{eqnarray}
In this case, it is easy to see via Eqs.\ (\ref{expinvq})\footnote{
One should replace $C_2\alpha_s$ by $\alpha_{\rm QED}$.  } and
(\ref{Sommerfeldfac}) that the QED correction including the boundstate
effects is twice as large as the (naive) one-loop correction at
$\sqrt{Q^2}=2m_q$.  One may regard this as a reference for the
significance of boundstate effects, with respect to the naive one-loop
correction, in the limit where the running of the strong coupling
constant may be neglected (since it corresponds to the limit where the
static QCD potential reduces to the Coulomb potential); qualitatively
it is a reasonable approximation for final-state quarks as heavy as
the top quark.

With Eq.\ (\ref{Pi(resum)}), we also define the magnitude of the QCD
correction relative to the LO contribution as
\begin{eqnarray}
  R (Q^2) _{\rm NLO,\,NR}= 
  \frac{[ \Pi_{\rm S}^{(q)} (Q^2)  ]_{\rm NLO,\,NR}}
  {[ \Pi_{\rm S}^{(q)} (Q^2)  ]_{\rm tree}} - 1.
\end{eqnarray}

\subsection{QCD-corrected neutralino annihilation cross section}

In calculating the relic density of the lightest neutralino, the
annihilation cross section for $Q^2\sim 4m_{\chi^0_1}^2$ is important.
This is because the lightest neutralinos decouple from the thermal
bath when the cosmic temperature is much lower than $m_{\chi^0_1}$; at
such low temperature, neutralinos are non-relativistic.

\begin{figure}[t]
  \centerline{\epsfxsize=0.6\textwidth\epsfbox{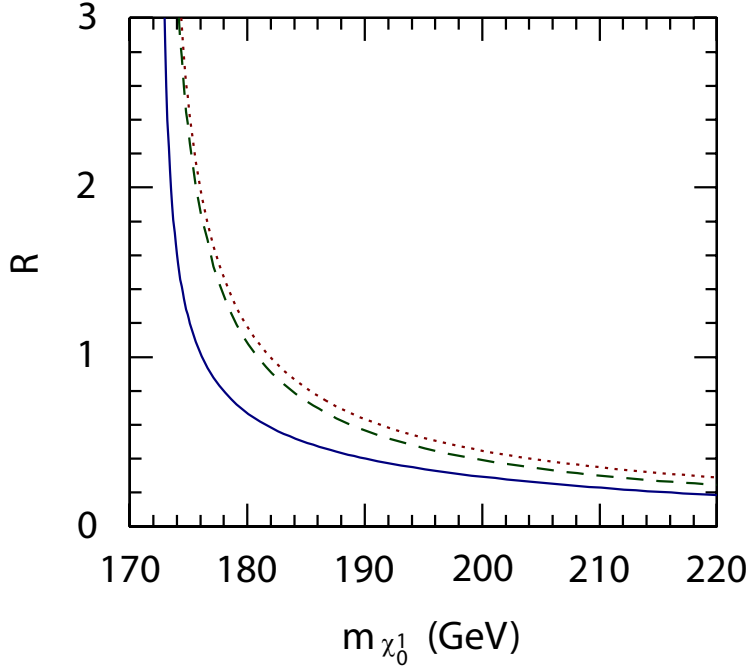}}
  \caption{QCD correction to the cross section
    $\sigma_{\chi^0_1\chi^0_1\rightarrow q\bar{q}(g)}$ when the
    initial $\chi^0_1$ are at rest (i.e., $Q^2\rightarrow
    4m_{\chi^0_1}^2$) as a function of the LSP mass $m_{\chi^0_1}$.
    Three lines correspond to $R_{O(\alpha_s)}$ (solid), $R_{\rm
    NLO,\,NR}$ with $\mu_{\rm Bohr}=20\ {\rm GeV}$ (dashed), and
    $R_{\rm NLO,\,NR}$ with $\mu_{\rm Bohr}=40\ {\rm GeV}$ (dotted).
    The input values are taken as $\alpha_s(m_t)=0.1080$ (corresponding
    to $\alpha_s(m_Z)=0.1187$) and 
    $m_t=172.5\ {\rm GeV}$.
    $\mu_{\rm Bohr}$ represents the renormalization scale in 
   $V_{\rm QCD}^{\rm (NLO)}(r)$, where
   a sensible choice of $\mu_{\rm Bohr}$ is at the Bohr scale of
   $t\bar{t}$ resonances.
   We may regard the difference between the predictions for
   $\mu_{\rm Bohr}=20$~GeV and 40~GeV as an estimate of contributions
   beyond NLO.
   }
  \label{fig:csratio}
\end{figure}

To see the importance of the QCD correction in such a case, in Fig.\
\ref{fig:csratio}, we plot $R_{O(\alpha_s)}(Q^2=4m_{\chi^0_1}^2)$ and
$R_{\rm NLO,\,NR}(Q^2=4m_{\chi^0_1}^2)$ as functions of
$m_{\chi^0_1}$.  It can be easily seen that the QCD correction
significantly enhances the cross section.  Since the tree-level cross
section vanishes as $v_t\rightarrow 0$ while the QCD corrections stay
constant, $R_{O(\alpha_s)}(Q^2=4m_{\chi^0_1}^2)$ and $R_{\rm
NLO,\,NR}(Q^2=4m_{\chi^0_1}^2)$ diverge as $m_{\chi^0_1}\rightarrow
m_t$.  As a result, the size of the QCD correction becomes comparable
to the tree-level contribution when the final-state top and anti-top
quarks are non-relativistic.  This fact has important implications to
the calculation of $\Omega_{\chi^0_1}$ when the lightest neutralino
dominantly annihilates into a $t\bar{t}$ pair.

Estimate of the annihilation cross section using the one-loop QCD
correction $R_{O(\alpha_s)}$ is valid when $\alpha_s\ll v_t$ and
$\alpha_s \ll 1$, since both (naive) higher-order contributions
$\alpha_s^n$ and higher-order boundstate contributions
$(\alpha_s/v_t)^n$ are suppressed.  On the other hand, estimate of the
QCD correction by $R_{\rm NLO,\,NR}$ is valid when $v_t, \, \alpha_s
\ll 1$.  Thus, in the overlap region when $\alpha_s\ll v_t \ll 1$
hold, both $R_{O(\alpha_s)}$ and $R_{\rm NLO,\,NR}$ are reasonable
approximations.  According to Fig.~\ref{fig:csratio}, when the
lightest neutralino mass is in the range $200-220\ {\rm GeV}$,
$R_{O(\alpha_s)}(Q^2=4m_{\chi^0_1}^2)$ is reasonably
close to $R_{\rm NLO, \, NR}(Q^2=4m_{\chi^0_1}^2)$
(with respect to the estimate of the NNLO correction).  
On the contrary, with smaller
$m_{\chi^0_1}$, boundstate effects become important.  Indeed, in such
a case, $R_{O(\alpha_s)}(Q^2=4m_{\chi^0_1}^2)$ and $R_{\rm
NLO,\,NR}(Q^2=4m_{\chi^0_1}^2)$ differ significantly, and the latter
prediction is more reliable.

Since the QCD correction enhances the cross section
$\sigma_{\chi^0_1\chi^0_1\rightarrow t\bar{t}(g)}$, it may play an
important role in the calculation of the relic density of
$\chi^0_1$. In the next section, we will discuss how the relic density
changes as we take account of the QCD correction.

%\subsection{Error estimates}

\section{Relic Density of the Lightest Neutralino}
\setcounter{equation}{0}
\label{sec:Omega}

In this section, we calculate the relic density of the lightest
neutralino.  We include the QCD correction to the neutralino
annihilation cross section given in the previous section and see its
effects on the calculation of the relic density of the neutralino.

QCD corrections for the process $\chi^0_1\chi^0_1\rightarrow
q\bar{q}(g)$ may be important for the precise calculation of
$\Omega_{\chi^0_1}$.  When the MSSM parameters will be precisely
measured in future collider experiments after the discoveries of
superparticles, cross sections of the pair annihilation processes of
$\chi^0_1$ will be calculated in detail.  Using the formulas given in
the previous section, it is straightforward to include the QCD
correction to the processes $\chi^0_1\chi^0_1\rightarrow q\bar{q}(g)$.
Here, we show the importance of the QCD correction.  To make our point
clearer, here we take into account only the most important
annihilation processes and neglect sub-dominant ones.

In order to calculate the relic density, we should follow the
evolution of the number density of the lightest neutralino
$n_{\chi^0_1}$, which is governed by the Boltzmann equation.  Denoting
the energy distribution function of the lightest neutralino as
$f_{\chi^0_1}$,\footnote{$n_{\chi^0_1}$ and $f_{\chi^0_1}$ are related
to each other by $n_{\chi^0_1} = \int \frac{d^3
\vec{p}_{\chi^0_1}}{(2\pi)^3} f_{\chi^0_1}(p_{\chi^0_1})$.} the
Boltzmann equation is given by \cite{KolbTurner}
\begin{eqnarray}
  \frac{d n_{\chi^0_1}}{dt} + 3 H n_{\chi^0_1} &=&
  - \sum_{\rm processes} 
  \int d\Pi_{\chi^0_1} d\Pi'_{\chi^0_1} d\Phi_{\rm f}
  \left| {\cal M}_{\chi^0_1\chi^0_1\rightarrow {\rm f}} \right|^2 
  f_{\chi^0_1}(p_{\chi^0_1}) f_{\chi^0_1}(p'_{\chi^0_1})
  \nonumber \\ &&
  + ({\rm production\ term}),
  \label{BoltzmannEq}
\end{eqnarray}
where $H$ is the expansion rate of the universe.  In the above
formula, $\int d\Phi_{\rm f}$ is the integration over the phase space of
final-state particles and we sum over all the relevant pair
annihilation processes.  In addition, the phase space integrations
$\int d\Pi_{\chi^0_1}$ and $\int d\Pi'_{\chi^0_1}$ are for 
the initial-state
neutralinos (whose momenta are $p_{\chi^0_1}$ and $p'_{\chi^0_1}$,
respectively).

Even if the superparticles may not be in {\it chemical} equilibrium,
they are expected to be in {\it kinematical} equilibrium since
superparticles effectively interact with ordinary standard-model
particles in thermal bath.  In addition, since the freeze out of the
lightest neutralino occurs when $T\ll m_{\chi^0_1}$, we use the
Boltzmann distribution function for neutralinos; at cosmic temperature
$T$,
\begin{eqnarray}
  f_{\chi^0_1}(p_{\chi^0_1}) = 2 e^{-(p_{\chi^0_1,0}-\mu_{\chi^0_1})/T},
  \label{f_X(kin.eq.)}
\end{eqnarray}
where the factor of $2$ represents the spin multiplicity of neutralino,
$p_{\chi^0_1,0}$ is the energy of $\chi^0_1$, and $\mu_{\chi^0_1}$ is
the chemical potential.

In order to follow the evolution of $n_{\chi^0_1}$ for $T\ll
m_{\chi^0_1}$, it is convenient to define
\begin{eqnarray}
  x(T) = \frac{T}{m_{\chi^0_1}}.
\end{eqnarray}
Once the temperature becomes lower than $m_{\chi^0_1}$, $x$ can be
used as an expansion parameter; we expand various quantities using
$x$.  For example, the number density of the lightest neutralino is
given by
\begin{eqnarray}
  n_{\chi^0_1} = 2 e^{-(m_{\chi^0_1}-\mu_{\chi^0_1})/T}
  \left( \frac{m_{\chi^0_1}T}{2\pi} \right)^{3/2}
  \left[
    1 - \frac{15}{8} x + O(x^2)
  \right],
  \label{n_X(NR)}
\end{eqnarray}
where this formula is obtained by expanding the energy as
\begin{eqnarray*}
  p_{\chi^0_1,0} = m_{\chi^0_1} + \frac{|\vec{p}_{\chi^0_1}|^2}{2m_{\chi^0_1}}
  - \frac{|\vec{p}_{\chi^0_1}|^4}{8m_{\chi^0_1}^3} + \cdots,
\end{eqnarray*}
with $\vec{p}_{\chi^0_1}$ being the three-momentum of $\chi^0_1$.
Substituting Eq.\ (\ref{f_X(kin.eq.)}) into Eq.\ (\ref{BoltzmannEq}),
and applying the detailed balance theorem to determine the production
term, the Boltzmann equation reduces to
\begin{eqnarray}
  \frac{d n_{\chi^0_1}}{dt} + 3 H n_{\chi^0_1} = 
  - \left( 1 - \frac{3}{2} x \right)
  {\cal S}_{\rm tot} (Q^2=4m_{\chi^0_1}^2 + 6 x m_{\chi^0_1}^2)
  \left( n_{\chi^0_1}^2 - n_{\chi^0_1}^{{\rm (eq)} 2}\right)
  +O(x^2),
  \label{BoltzmannEqNR}
\end{eqnarray}
where $n_{\chi^0_1}^{{\rm (eq)}}$ is the chemical-equilibrium value of
the number density of $\chi^0_1$, which is obtained from Eq.\
(\ref{n_X(NR)}) by taking $\mu_{\chi^0_1}\rightarrow 0$.  In
addition, ${\cal S}_{\rm tot}$ is given by
\begin{eqnarray}
  {\cal S}_{\rm tot} = 2 v_\chi \sigma_{\chi^0_1\chi^0_1\rightarrow {\rm all}},
\end{eqnarray}
with $\sigma_{\chi^0_1\chi^0_1\rightarrow {\rm all}}$ being the total
pair annihilation cross section of the lightest
neutralino.\footnote{Here, we assume that the effects of
coannihilation is negligible.  For a precise calculation of the relic
density, effects of the coannihilation should be taken into account
\cite{Allanach:2004xn}.  Once the relevant MSSM parameters are
measured at the future collider experiments, however, effects of the
coannihilation can be theoretically controlled.  Thus, in our
analysis, we do not include effects of the coannihilation,
which are beyond our scope.}  Taking
account of the correction of the order of $x$, $Q^2$ is shifted from
$4m_{\chi^0_1}^2$ to $4m_{\chi^0_1}^2+6xm_{\chi^0_1}^2$.  This is due
to the fact that, in the thermal bath, neutralinos have non-vanishing
kinetic energy whose average is $\frac{3}{2}T$.

From Eq.\ (\ref{BoltzmannEqNR}), several important features of
$n_{\chi^0_1}$ are derived.  When the cosmic temperature is high
enough, production and annihilation rates of the lightest neutralino,
which are $\sim {\cal S}_{\rm tot}n_{\chi^0_1}$, are relatively high.
At this epoch, scattering rate becomes much larger than the cosmic
expansion rate $H$ and, consequently, the number density of $\chi^0_1$
keeps up with the equilibrium value $n_{\chi^0_1}^{\rm (eq)}$.  As the
temperature decreases, however, the situation changes.  Importantly,
once the cosmic temperature becomes lower than $m_{\chi^0_1}$,
$n_{\chi^0_1}^{\rm (eq)}$ is Boltzmann suppressed.  When
$n_{\chi^0_1}$ (which is close to $n_{\chi^0_1}^{\rm (eq)}$ at this
epoch) becomes so suppressed that ${\cal S}_{\rm tot}n_{\chi^0_1}$
becomes smaller than $H$, $\chi^0_1$ freezes out from the thermal
bath. After the freeze out, the number of $\chi^0_1$ in comoving volume is
conserved.

The freeze-out of the lightest neutralino occurs when $x^{-1}\sim 20$
and hence the lightest neutralinos are non-relativistic in thermal
bath at the time of freeze out.  Thus, the relic density is primarily
determined by the behavior of the total cross section at
$Q^2\rightarrow 4m_{\chi^0_1}^2$, as can be understood from Eq.\
(\ref{BoltzmannEqNR}).  When $A_\chi$ given in Eq.\ (\ref{A_chi}) is
sizable, pair annihilation of the lightest neutralino is dominated by
$s$-wave processes as $Q^2\rightarrow 4m_{\chi^0_1}^2$.  Thus, in
such a case, $\Omega_{\chi^0_1}$ is mostly determined by the $s$-wave
part of the total cross section.  When all the sfermions and heavier
Higgses are relatively heavy, the relevant $s$-wave processes are
$\chi^0_1\chi^0_1\rightarrow t\bar{t}(g)$, $W^+W^-$, and $ZZ$.  In our
analysis, we assume that this is the case and approximate
\begin{eqnarray}
 {\cal S}_{\rm tot} \simeq
 {\cal S}_{\chi^0_1\chi^0_1\rightarrow Z^*\rightarrow q\bar{q}(g)} + 
 {\cal S}_{\chi^0_1\chi^0_1\rightarrow W^+W^-} +
 {\cal S}_{\chi^0_1\chi^0_1\rightarrow ZZ}.
\end{eqnarray}
Here, the second and third terms on the right-hand side represent the
contributions from $W^+W^-$ and $ZZ$ final states, respectively:
\begin{eqnarray}
 {\cal S}_{\chi^0_1\chi^0_1\rightarrow W^+W^-, ZZ} = 
 2 v_{\chi^0_1} \sigma_{\chi^0_1\chi^0_1\rightarrow W^+W^-, ZZ},
\end{eqnarray}
with $\sigma_{\chi^0_1\chi^0_1\rightarrow W^+W^-, ZZ}$ being total
cross section for the corresponding processes.  We use 
the tree-level formulas
for $\sigma_{\chi^0_1\chi^0_1\rightarrow W^+W^-}$ and
$\sigma_{\chi^0_1\chi^0_1\rightarrow ZZ}$.  In addition, we evaluate
these quantities at the threshold since their $Q^2$ dependences are
weak when $x\ll 1$.  (For $\chi^0_1\chi^0_1\rightarrow q\bar{q}(g)$,
$O(x)$ terms are considered unless otherwise mentioned.)  Other
processes, like $\chi^0_1\chi^0_1\rightarrow Zh$ and $hh$, are
$p$-wave processes and are neglected in our study.  If we include
those processes, $\Omega_{\chi^0_1}$ may decrease by a few $\%$ or so
\cite{Allanach:2004xn}; once the MSSM parameters will be
experimentally determined, it will be straightforward to take account
of effects of the $p$-wave processes.  Furthermore, ${\cal
S}_{\chi^0_1\chi^0_1\rightarrow q\bar{q}(g)}$ is calculated 
neglecting the $h$-exchange diagram since its contribution is $p$-wave
suppressed.

We have numerically solved the Boltzmann equation
(\ref{BoltzmannEqNR}) to determine the present value of
$n_{\chi^0_1}$.  We impose the initial condition
$n_{\chi^0_1}=n_{\chi^0_1}^{\rm (eq)}$ at the cosmic time where ${\cal
  S}_{\rm tot}n_{\chi^0_1}^{\rm (eq)}\gg H$ and follow the evolution
of $n_{\chi^0_1}$.  Then, we determine the number density of the
lightest neutralino at ${\cal S}_{\rm tot}n_{\chi^0_1}^{\rm (eq)}\ll
H$.  In our following argument, it is convenient to define the yield
variable
\begin{eqnarray}
  Y_{\chi^0_1} = \frac{n_{\chi^0_1}}{s},
\end{eqnarray}
with $s$ being the entropy density.  Important property of this quantity
is that $Y_{\chi^0_1}$ stays constant when ${\cal S}_{\rm
  tot}n_{\chi^0_1}^{\rm (eq)}\ll H$; indeed, we have numerically
checked that the yield variable becomes a constant of time when the
scattering rate is much smaller than the expansion rate.  Using the
yield variable after the freeze-out of $\chi^0_1$ (which we denote
$Y^{({\rm now})}_{\chi^0_1}$), the density parameter is given by
\begin{eqnarray}
  \Omega_{\chi^0_1} = 
  m_{\chi^0_1} Y^{({\rm now})}_{\chi^0_1} 
  \left[ \frac{\rho_{\rm crit}}{s^{({\rm now})}} \right]^{-1}.
\end{eqnarray}
Here, $\rho_{\rm crit}$ is the critical density and $s^{({\rm now})}$
is the present entropy density; numerically,
\begin{eqnarray}
  \frac{\rho_{\rm crit}}{s^{({\rm now})}} \simeq
  3.6 \times 10^{-9} h_{100}^2\ {\rm GeV}.
\end{eqnarray}

\begin{figure}
  \centerline{\epsfxsize=0.6\textwidth\epsfbox{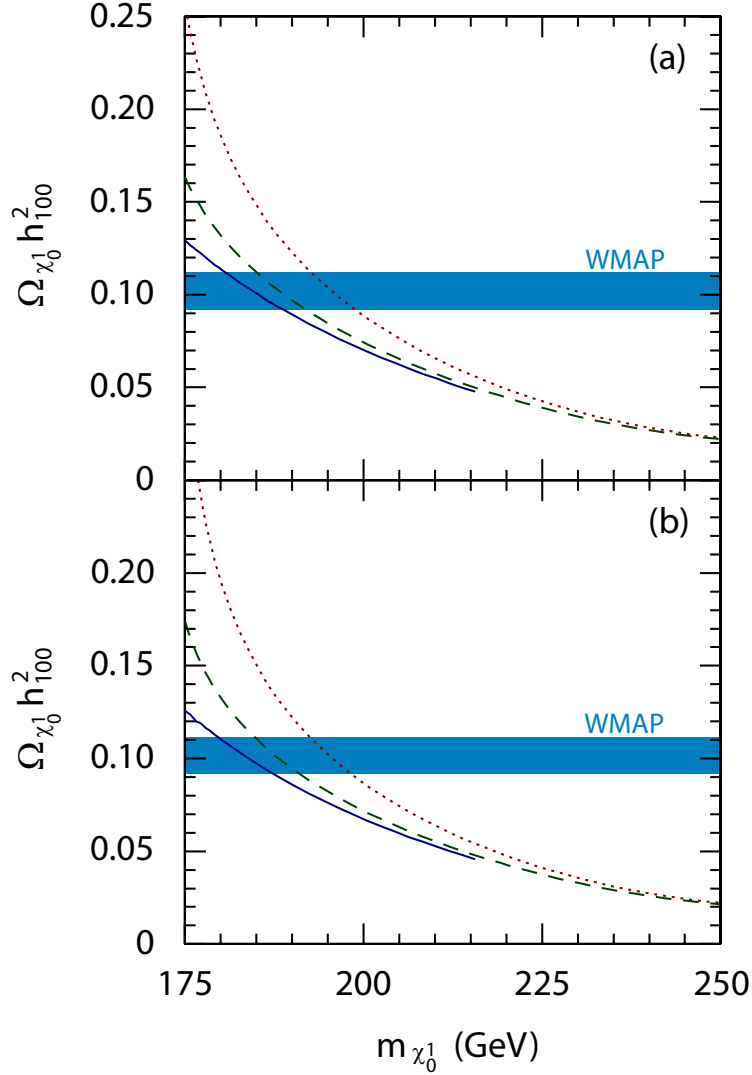}}
  \caption{Density parameter of the lightest neutralino as a function
    of $m_{\chi^0_1}$.  The cross section
    $\sigma_{\chi^0_1\chi^0_1\rightarrow t\bar{t}(g)}$ is evaluated
    with NLO QCD correction with boundstate effects (solid),
    $O(\alpha_s)$ correction (dashed), and tree-level formula
    (dotted).  Here we take $\alpha_s(m_Z)=0.1189$, $m_t=172.5\ {\rm
    GeV}$, and $\mu_H=280\ {\rm GeV}$.  The dark shaded band shows the
    dark matter density measured by the WMAP (see Eq.\
    (\ref{Omega_c(WMAP)})).  For (a) and (b), collision term for
    $t\bar{t}(g)$ final state is evaluated up to $O(x)$ and $O(x^0)$,
    respectively.}
  \label{fig:omg280}
\end{figure}

In Fig.\ \ref{fig:omg280}, we plot the density parameter of the
lightest neutralino as a function of $m_{\chi^0_1}$.  In the figure,
we compare the results with and without the QCD
correction to the process $\chi^0_1\chi^0_1\rightarrow t\bar{t}(g)$;
we evaluate the cross section $\sigma_{\chi^0_1\chi^0_1\rightarrow
t\bar{t}(g)}$ with NLO QCD correction with boundstate effects (solid),
$O(\alpha_s)$ correction (dashed), and tree-level formula (dotted).
In the calculation with the boundstate effects, we restrict ourselves
to the case with $m_{\chi^0_1}\lesssim 210\ {\rm GeV}$.  This is
because, when $m_{\chi^0_1}$ is too large, 
non-relativistic approximation used in
the calculation becomes unreliable.  Fortunately, however, as the
lightest neutralino mass becomes larger than $\sim 200\ {\rm
GeV}$, result with the boundstate effects is in a very good agreement
with that with $O(\alpha_s)$ contributions.  This fact shows that QCD
correction to the process $\chi^0_1\chi^0_1\rightarrow t\bar{t}(g)$ is
well approximated by the one-loop formula when the lightest neutralino
mass is larger than $\sim 200\ {\rm GeV}$.  On the contrary, when
$m_{\chi^0_1}$ is smaller, difference between the one-loop result and
the resummed one is quite large; in this case the latter prediction
(solid) is more reliable.

We can see that the QCD correction significantly reduces the density
parameter of the lightest neutralino when $m_{\chi^0_1}$ is close to
the top-quark mass.  Since the $t\bar{t}$ final state is more
important than the gauge-boson final states when the lightest
neutralino mass is relatively close to the top-quark mass, this fact
can be easily understood from Fig.\ \ref{fig:csratio}.  As
$m_{\chi^0_1}$ becomes larger, on the contrary,
$\sigma_{\chi^0_1\chi^0_1\rightarrow W^+W^-, ZZ}$ becomes larger than
$\sigma_{\chi^0_1\chi^0_1\rightarrow t\bar{t}(g)}$ and the gauge-boson
final state becomes more important than the $t\bar{t}$ final state.  In
such a case, even if the QCD correction to the process
$\chi^0_1\chi^0_1\rightarrow t\bar{t}(g)$ is sizable,
$\Omega_{\chi^0_1}$ may not be affected so much.  Indeed, we can see
such a behavior in the figure.  As the lightest neutralino mass
becomes large enough, tree-level result becomes almost the same as the
one with QCD correction although the QCD correction to
$\sigma_{\chi^0_1\chi^0_1\rightarrow t\bar{t}(g)}$ is still more than
$10\ \%$.

In Fig.\ \ref{fig:omg280}(a), we kept $O(x)$ terms of ${\cal
S}_{\chi^0_1\chi^0_1\rightarrow t\bar{t}(g)}$.  For comparison, we
also calculated the density parameter with neglecting $x$-dependences
(by taking $x\rightarrow 0$); the results are shown in Fig.\
\ref{fig:omg280}(b).  We can see that the resultant density parameters
differ in two figures when the lightest neutralino mass becomes close
to the top-quark mass.  We have checked that the effects of $O(x)$ terms
become larger as $m_{\chi^0_1}$ becomes closer to $m_t$.

\begin{figure}
  \centerline{\epsfxsize=0.6\textwidth\epsfbox{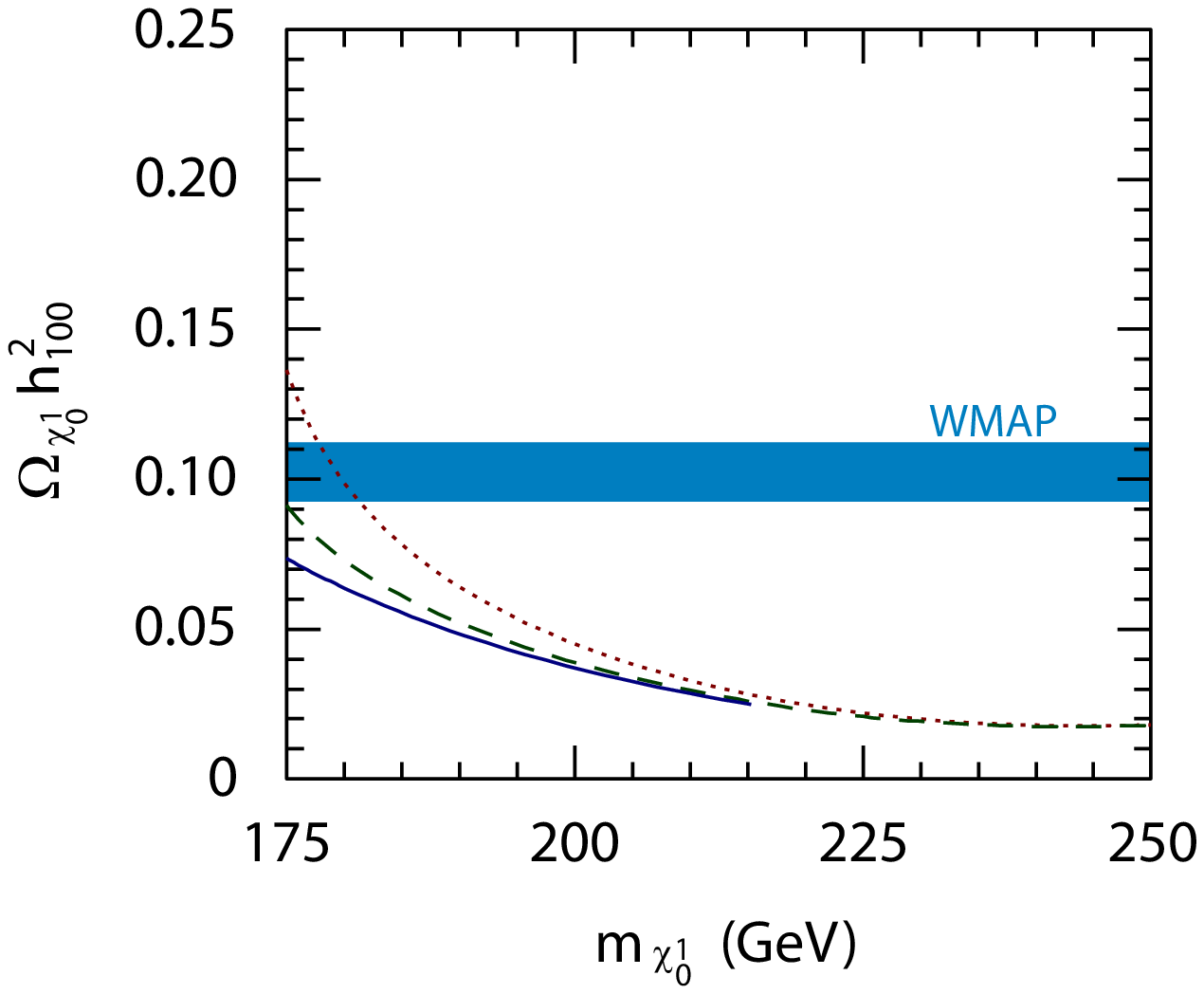}}
  \caption{Same as Fig.\ \ref{fig:omg280}(a) except for $\mu_H=260\ {\rm
      GeV}$.}
  \label{fig:omg260}
\end{figure}

\begin{figure}
  \centerline{\epsfxsize=0.6\textwidth\epsfbox{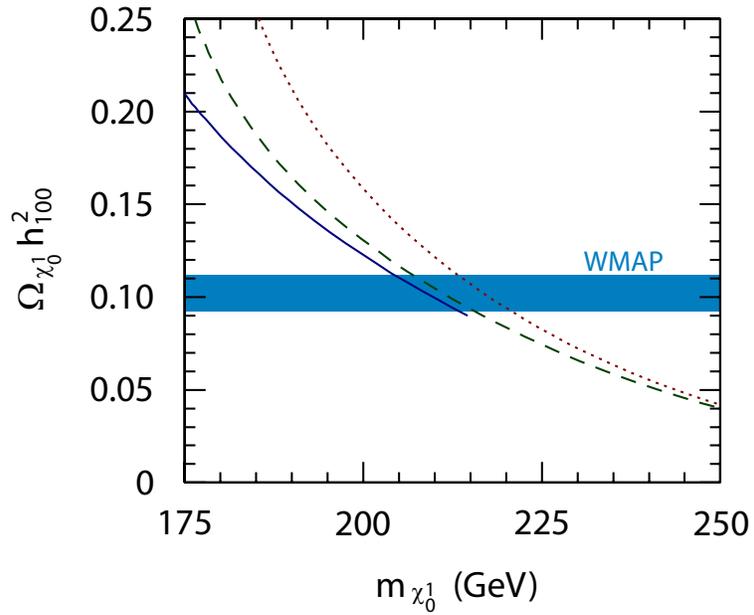}}
  \caption{Same as Fig.\ \ref{fig:omg280}(a) except for $\mu_H=300\ {\rm
      GeV}$.}
  \label{fig:omg300}
\end{figure}

In Figs.\ \ref{fig:omg260} and \ref{fig:omg300}, we also plot the
density parameter for the cases with $\mu_H=260\ {\rm GeV}$ and
$\mu_H=300\ {\rm GeV}$, respectively.  As we increase $\mu_H$, the
WMAP value is realized with larger value of $m_{\chi^0_1}$ due to the
change of the mixing matrix of the neutralinos.  These figures show
that, with these values of $\mu_H$, the relic density of the lightest
neutralino is significantly affected when $m_{\chi^0_1}$ is close to
$m_t$.  As $\mu_H$ increases, however, QCD correction becomes less
important in the parameter region where $\Omega_{\chi^0_1}$ agrees
with the WMAP value.

\section{Conclusions and Discussion}
\setcounter{equation}{0}
\label{sec:conclusion}

In this paper, we have calculated QCD correction to the process
$\chi^0_1\chi^0_1\rightarrow q\bar{q}(g)$, and discussed its
implications to the determination of $\Omega_{\chi^0_1}$.  We have
seen that the QCD correction enhances the cross section
$\sigma_{\chi^0_1\chi^0_1\rightarrow q\bar{q}(g)}$ by $O(10\ \%)$ when
the final-state quarks are non-relativistic.  Importantly, the
lightest neutralino dominantly annihilates into a $t\bar{t}$ pair (as
well as into $W^+W^-$ and $ZZ$) when the lightest neutralino has
sizable Higgsino components.  In such a case, QCD correction may
become large and significantly enhance the total annihilation cross
section of the lightest neutralino.  Since the QCD correction is
enhanced close to the threshold by factors $(\alpha_s/v_t)^k$, the
QCD correction becomes important in particular when $m_{\chi^0_1}$ is
close to the top-quark mass.  Since the relic density of the lightest
neutralino is (approximately) inversely proportional to the total pair
annihilation cross section, the relic density may be reduced by $O(10\
\%)$ by the QCD correction.  Thus, for a precise calculation of the
relic density of the LSP, QCD correction should be taken into account
when $\chi^0_1$ dominantly annihilates into the $t\bar{t}$ final state.

Since the dark matter density has been determined with an accuracy of
$O( 10\ \%)$ by WMAP and other experiments, this fact has important
implications for future studies of supersymmetric dark matter(s).
Once properties of superparticles are determined in future collider
experiments, it will become possible to theoretically calculate the
relic density of the lightest neutralino (if it is the LSP).
Comparison of the result of such a calculation with the observed
density of the dark matter will be a crucial test of the scenario of
the lightest-neutralino dark matter.  In addition, even if the
lightest neutralino is {\it not} the LSP, precise determination of the
freeze-out density of the lightest neutralino may be important for
understanding the origin of the dark matter.  In particular,
possibilities of very weakly interacting LSP (like gravitino-, axino-,
or right-handed-sneutrino-LSP), which are called superWIMPs, have been
intensively considered recently.  If one of the superWIMPs is the LSP,
the lightest neutralino becomes unstable.  In this case, however, the
lifetime of the lightest neutralino is usually very long and hence, if
$\chi^0_1$ is the lightest superparticle in the MSSM sector, thermally
produced lightest neutralinos decay after they freeze out from the
thermal bath.  If this is the case, thermally produced lightest
neutralino becomes one of the significant sources of the superWIMP
cold dark matter.  Thus, even in the scenario with superWIMPs, it is
important to calculate the relic density of the lightest neutralino.
\medbreak

\noindent
{\it Acknowledgements}: The authors would like to thank
T. Asaka, K. Hikasa and M. Tanabashi for useful discussion and
valuable comments.


\begin{thebibliography}{99}

\bibitem{WMAP2006}
  WMAP webpage, 
  {\tt http://lambda.gsfc.nasa.gov}.

\bibitem{CDMatLC}
  T.~Moroi, Y.~Shimizu and A.~Yotsuyanagi,
  Phys.\ Lett.\ B {\bf 625}, 79 (2005);
  %%CITATION = HEP-PH 0505252;%%
  T.~Moroi and Y.~Shimizu,
  Phys.\ Rev.\ D {\bf 72}, 115012 (2005);
  %%CITATION = HEP-PH 0509196;%%
  E.~A.~Baltz, M.~Battaglia, M.~E.~Peskin and T.~Wizansky,
  arXiv:hep-ph/0602187.
  %%CITATION = HEP-PH 0602187;%%

\bibitem{LSPDM_WMAP}
  J.~R.~Ellis, K.~A.~Olive, Y.~Santoso and V.~C.~Spanos,
  Phys.\ Lett.\ B {\bf 565}, 176 (2003);
  %%CITATION = HEP-PH 0303043;%%
  H.~Baer and C.~Balazs,
  JCAP {\bf 0305}, 006 (2003);
  %%CITATION = HEP-PH 0303114;%%
  U.~Chattopadhyay, A.~Corsetti and P.~Nath,
  Phys.\ Rev.\ D {\bf 68}, 035005 (2003).
  %%CITATION = HEP-PH 0303201;%%

\bibitem{Barger:2005ve}
  V.~Barger, W.~Y.~Keung, H.~E.~Logan, G.~Shaughnessy and A.~Tregre,
  Phys.\ Lett.\ B {\bf 633}, 98 (2006).
  %%CITATION = HEP-PH 0510257;%%

\bibitem{focus}
  J.~L.~Feng and T.~Moroi,
  Phys.\ Rev.\ D {\bf 61}, 095004 (2000);
  %% CITATION = HEP-PH 9907319;%%
  J.~L.~Feng, K.~T.~Matchev and T.~Moroi,
  Phys.\ Rev.\ Lett.\  {\bf 84}, 2322 (2000);
  %% CITATION = HEP-PH 9908309;%%
  Phys.\ Rev.\ D {\bf 61}, 075005 (2000).
  %% CITATION = HEP-PH 9909334;%%

\bibitem{Feng:2000gh}
  J.~L.~Feng, K.~T.~Matchev and F.~Wilczek,
  Phys.\ Lett.\ B {\bf 482}, 388 (2000).
  %% CITATION = HEP-PH 0004043;%%
  
\bibitem{Ibe:2005jf}
  M.~Ibe, T.~Moroi and T.~Yanagida,
  Phys.\ Lett.\ B {\bf 620}, 9 (2005).
  %% CITATION = HEP-PH 0502074;%%

\bibitem{Hoang:2000yr}
  See, for example,
  A.~H.~Hoang {\it et al.},
  Eur.\ Phys.\ J.\ directC {\bf 2}, 1 (2000),
  %%CITATION = HEP-PH 0001286;%%
  and references therein.
  
\bibitem{GravitinoCDM}
  T.~Moroi, H.~Murayama and M.~Yamaguchi,
  Phys.\ Lett.\ B {\bf 303}, 289 (1993);
  %%CITATION = PHLTA,B303,289;%%
  J.~L.~Feng, A.~Rajaraman and F.~Takayama,
  Phys.\ Rev.\ Lett.\  {\bf 91}, 011302 (2003).
  %%CITATION = HEP-PH 0302215;%%

\bibitem{AxinoCDM}
  L.~Covi, J.~E.~Kim and L.~Roszkowski,
  Phys.\ Rev.\ Lett.\  {\bf 82}, 4180 (1999).
  %%CITATION = HEP-PH 9905212;%%

\bibitem{Asaka:2005cn}
  T.~Asaka, K.~Ishiwata and T.~Moroi,
  Phys.\ Rev.\ D {\bf 73}, 051301 (2006).
  %%CITATION = HEP-PH 0512118;%%

\bibitem{Gunion:1989we}
  J.~F.~Gunion, H.~E.~Haber, G.~L.~Kane and S.~Dawson,
  ``The Higgs Hunter's Guide,'' (1990, Addison-Wesley Publishing).

\bibitem{Hikasa:1996wp}
  K.~i.~Hikasa and J.~Hisano,
  Phys.\ Rev.\ D {\bf 54}, 1908 (1996).
  %%CITATION = HEP-PH 9603203;%%

\bibitem{Drees:1990dq}
  M.~Drees and K.~i.~Hikasa,
  Phys.\ Lett.\ B {\bf 240}, 455 (1990)
  [Erratum-ibid.\ B {\bf 262}, 497 (1991)].
  %%CITATION = PHLTA,B240,455;%%

\bibitem{Fischler:1977yf}
  W.~Fischler,
  Nucl.\ Phys.\ B {\bf 129}, 157 (1977).
  %%CITATION = NUPHA,B129,157;%%

\bibitem{KolbTurner}
  E.~W.~Kolb and M.~S.~Turner,
  ``The Early Universe,'' (1990, Addison-Wesley Publishing).
  
\bibitem{Allanach:2004xn}
  B.~C.~Allanach, G.~Belanger, F.~Boudjema and A.~Pukhov,
  JHEP {\bf 0412}, 020 (2004).
  %%CITATION = HEP-PH 0410091;%%

\end{thebibliography}
\end{document}